# Phase-Stretch Adaptive Gradient-Field Extractor (PAGE)

Callen MacPhee, Madhuri Suthar, Bahram Jalali
Electrical and Computer Engineering Department
University of California, Los Angeles

*Abstract*— Phase-Stretch Adaptive Gradient-Field Extractor (PAGE) is an edge detection algorithm that is inspired by physics of electromagnetic diffraction and dispersion. A computational imaging algorithm, it identifies edges, their orientations and sharpness in a digital image where the image brightness changes abruptly. Edge detection is a basic operation performed by the eye and is crucial to visual perception. PAGE embeds an original image into a set of feature maps that can be used for object representation and classification. The algorithm performs exceptionally well as an edge and texture extractor in low light level and low contrast images. This manuscript is prepared to support the open-source code which is being simultaneously made available within the GitHub repository https://github.com/JalaliLabUCLA/Phase-Stretch-Adaptive-Gradient-field-Extractor/.

## I. OPERATIONAL PRINCIPAL

Phase-stretch Adaptive Gradient-field Extractor (PAGE) is a physics inspired feature engineering algorithm that computes a feature set comprised of edges at different spatial frequencies (and hence spatial scales) and orientations [1]. Metaphorically speaking, PAGE emulates the physics of birefringent (orientation-dependent) diffractive propagation through a physical medium with a specific diffractive structure. The propagation converts a real-valued image into a complex function. Related information is contained in the real and imaginary components of the output. The output represents the phase of the complex function. PAGE builds on the Phase stretch transform (PST) [2], another physics-inspired edge detection algorithm introduced by our laboratory. The Phase stretch transform algorithm evolved from the research on a class of real time measurement and sensing methods known as the photonic times stretch including time stretch analog-to-digital converter, time stretch dispersive Fourier transform and serial time-encoded amplified microscopy [4, 5, 6]. This manuscript is being released concurrently with the publishing of the code for PAGE on GitHub.

In a birefringent optical medium, the dielectric constant of the medium and hence, its refractive index is a function of spatial frequency and the polarization in the transverse plane. To understand the analogy between PAGE and electromagnetic propagation equations, let's consider an optical field with two linearly orthogonal polarizations propagating through a medium.

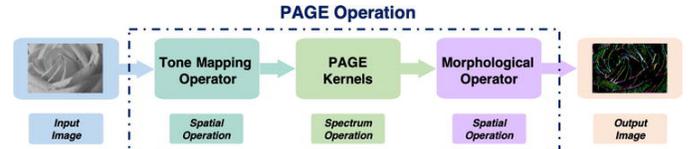

*Figure 1: Different steps of the phase-stretch gradient-field extractor (PAGE) algorithm.*

The Fourier content of the incoming signal,

$$\tilde{E}_i(u,v;z) = FFT\{E_i(x,y;z)\} \quad (1)$$

can be decomposed into the two orthogonal polarizations as

$$\tilde{E}_i(z) = \tilde{E}_x(z) + \tilde{E}_y(z) \quad (2)$$

where $FFT\{\}$ is the fast Fourier transform over the transversal coordinates and $(u,v)$ are spatial frequency variables.

As the propagation constant $\beta = \frac{2\pi n}{\lambda}$ is a function of refractive index, the two orthogonal polarizations $\tilde{E}_x$ and $\tilde{E}_y$ will have different propagation constants and hence, a phase difference at the output given by the following equation:

$$\Delta\phi = \phi_x - \phi_y = \Delta\beta = \frac{\omega_m}{c}|n_x - n_y| \quad (3)$$

By controlling the value of $n_x$ and $n_y$, as well the dependence of refractive index on frequency $n_x(\omega)$ and $n_y(\omega)$, coherent detection at the output detects a hyper-dimensional feature set from a 2D image that corresponds to edges at user-defined specific orientations and spatial frequencies. We note that in the above definition of the phase, we have set the propagation length to 1.

## II. MATHEMATICAL FOUNDATION

The first step is to apply an optional smoothening kernel in the frequency domain to reduce noise. This is typically performed in the frequency domain (after a Fourier transform), but it can also be done in the spatial domain using convolution. The image is then multiplied by a phase kernel that emulates the birefringence and frequency channelized diffractive propagation. Next, the image is transformed back into the spatial domain followed by a calculation of the spatial phase representing the desired feature vectors. The final step of PAGE is to apply thresholding and morphological operations on the generated feature vectors to produce the final output. For a color image, these operations are performed separately on all color channels and the results are then combined in a single image, although each channel can also be viewed separately.

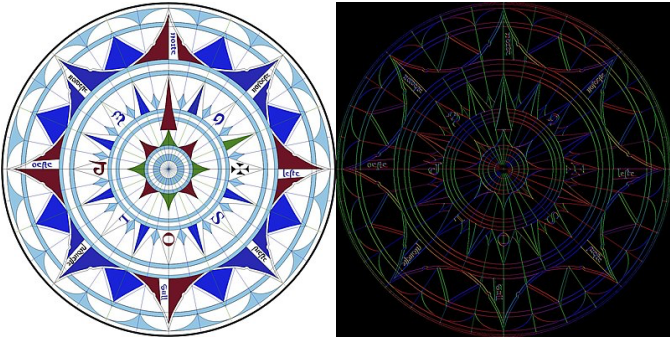

*Figure 2: (a) Phase-Stretch Adaptive Gradient-Field Extractor performed on a compass rose. The colors represent the orientation (angle) of the edge.*

Mathematically, this sequence of operations can be represented by the following equations. The birefringent Stretch operator $\mathbb{S}\{\}$ is defined as follows:

$$E_O[x,y] = \mathbb{S}\{E_i[x,y]\}$$
$$= IFFT\{\widetilde{K}[u,v,\theta] \cdot \widetilde{L}[u,v] \cdot FFT\{E_i[x,y]\}\} \quad (4)$$

where $E_O[x,y]$ is a complex quantity defined as,

$$E_O[x,y] = |E_O[x,y]|e^{-j\theta[x,y]} \quad (5)$$

In the above equations, $E_i[x,y]$ is the input image, $(x,y)$ are the spatial variables, the function $\widetilde{K}[u,v,\theta]$ is called the PAGE kernel and the function $\widetilde{L}[u,v]$ is a denoising kernel, both implemented in frequency domain. For the results shown, $\widetilde{L}[u,v]$ is a gaussian filter whose cut off frequency is determined by the sigma of the gaussian filter.

The PAGE operator $\mathbb{P}\{\}$ is then defined as the phase of the output of the stretch operation $\mathbb{S}\{\}$ applied on the input image $E_i[x,y]$:

$$\mathbb{P}\{E_i[x,y]\} = \measuredangle\{\mathbb{S}\{E_i[x,y]\}\} \quad (6)$$

where $\measuredangle\langle\cdot\rangle$ is the angle operator.

## III. PAGE FILTER BANKS

PAGE filter banks are defined by the PAGE kernel $\widetilde{K}[u,v,\theta]$ and are designed to compute semantic information from an image at different orientations and frequencies. The PAGE kernel $\widetilde{K}[u,v,\theta]$, consists of a phase filter which is a function of frequency variable $u$ and $v$, and the angle variable $\theta$ which controls the directionality of the edge. The spectral phase operator is expressed as a product of two phase functions, $\phi_1$ and $\phi_2$. The first component $\phi_1$ is a symmetric gaussian filter that selects the spatial frequency range of the edges that are detected. Default center frequency is 0, which indicates a baseband filter, the center frequency and bandwidth of which can be changed to probe edges with different sharpness. In other words, it enables the filtering of edges occurring over different spatial scales. The second component, $\phi_2$, performs the edge-detection. Since the output is based on the phase, it needs to be a complex-valued function. The PAGE operation transforms a real-value input to a complex-value quantity from which the phase is extracted.

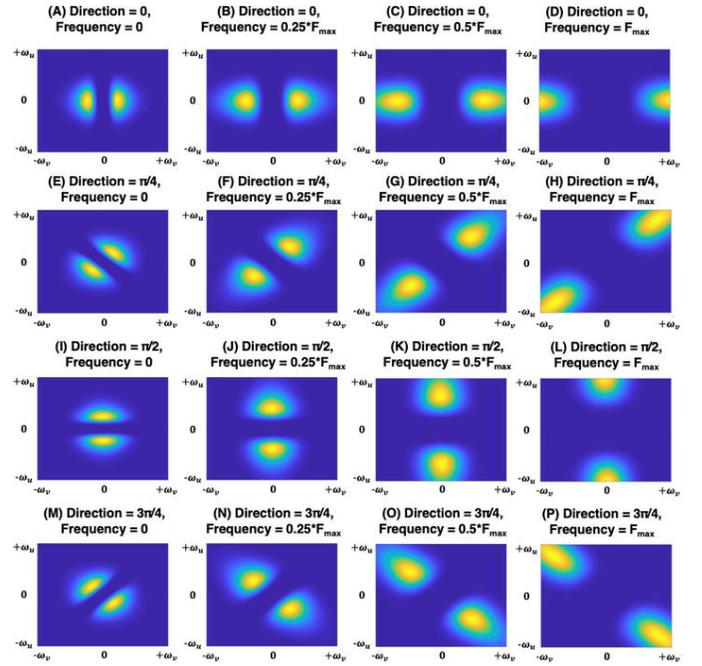

*Figure 3: Phase-stretch gradient-field extractor (PAGE) filter banks for various frequencies and directions.*

A change of basis leads to the transformed frequency variables $u'$ and $v'$

$$\begin{aligned} u' &= u \cdot \cos(\theta) + v \cdot \sin(\theta) \\ v' &= u \cdot \sin(\theta) + v \cdot \cos(\theta) \end{aligned} \quad (7)$$

such that the frequency vector rotates about the origin with $\theta$

$$u' + jv' \quad (8)$$

The PAGE kernel $\widetilde{K}[u,v,\theta]$ is defined as a function of frequency variable $(u,v)$ and angle $\theta$ as follows:

$$\widetilde{K}[u,v,\theta] = \widetilde{K}[u',v'] = exp\{j \cdot \phi_1(u') \cdot \phi_2(v')\} \quad (9)$$

where

$$\phi_1(u') = S_{u'} \cdot \frac{1}{\sqrt{2\pi}\sigma_{u'}} \cdot \exp\left(\frac{-(|u'| - \mu_{u'})^2}{2\sigma_{u'}^2}\right)$$
$$\phi_2(v') = S_{v'} \cdot \frac{1}{\sqrt{2\pi}|v'|\sigma_{v'}} \cdot \exp\left(\frac{-\ln(|v'| - \mu_{v'})^2}{2\sigma_{v'}^2}\right) \quad (10)$$

For all simulation examples here, the phase functions $\phi_1(u')$ and $\phi_2(v')$ are normalized in the range $(0,1)$ for all values of $\theta$ and then multiplied by $S_{u'}$ and $S_{v'}$ respectively, such that the strength of each kernel is mutable for different applications and image conditions.

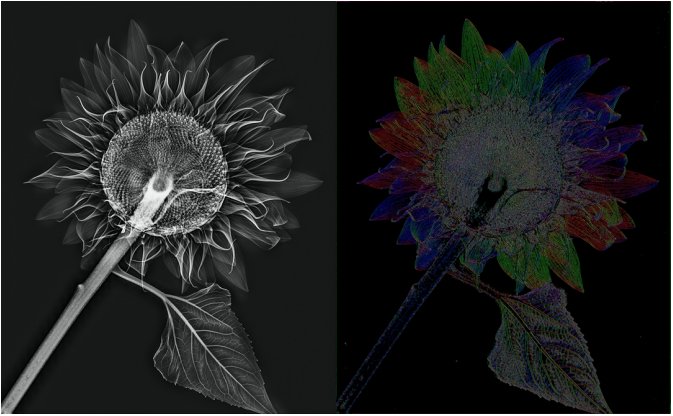

*Figure 4: Demonstration of the Phase-Stretch Adaptive Gradient-Field Extractor (PAGE) algorithm on an X-ray image of a sunflower. The colors represent the orientation (angle) of the edge.*

## IV. Feature Extraction and Application

PAGE has the potential to be used as a preprocessing step for machine learning tasks such as image classification. An important step in any classification task is feature extraction. In such applications, images are reduced to certain key features that aid in the tasks of object detection and classification. PAGE serves as a physics-inspired feature extractor and descriptor. It returns a hyper-dimensional feature mapping in which regions of great change in intensity are highlighted and grouped based on directionality. Given that it is selective over edge width and orientation, it is able to return a rich feature matrix with high representational power.

PAGE has a diverse set of applications that span several fields. Diagnosis and classification of retinopathy, for example, are medically important tasks highly dependent upon segmentation of blood vessels of varied width and orientation. This segmentation, and further image analysis, can be accomplished through a directional edge filter such as PAGE. Similarly, digital subtraction angiography creates an image of blood vessels using a contrast medium that can be used in pathology for soft tissue. Such imagery can be processed using the PAGE filter for diagnosis and visualization purposes. Further applications of note requiring directional edge information such as that computed by PAGE include fingerprint, written character, and flora and fauna recognition.

Originally introduced in 2020, PAGE builds on the Phase Stretch Transform (PST) [3]. Local Flow PST (LF-PST) is another algorithm introduced in 2020 that is based on PST and also performs orientation and scale dependent edge detection [7]. Local Flow PST has shown exceptional results in retina vessel detection for application to retinopathy.

# V. PAGE Algorithm Implementation

**Implementation of Phase-stretch Adaptive Gradient-Field Extractor (PAGE) in Python**
@contributors: Madhuri Suthar, Callen MacPhee, Yiming Zhou, Dale Capewell, Jalali Lab, Department of Electrical and Computer Engineering, UCLA — Additional Materials Located within https://github.com/JalaliLabUCLA/Phase-Stretch-Adaptive-Gradient-field-Extractor/.

PAGE or Phase-stretch Adaptive Gradient-field Extractor is a physics-inspired algorithm for detecting edges and their orientations in digital images at various scales. The algorithm is based on the diffraction equations of optics.

In the original implementation published in [1], the first step is to apply an adaptive tone mapping operator (TMO) to enhance the local contrast. Next, we reduce the noise by applying a smoothening kernel in frequency domain (this operation can also be done in spatial domain). We then apply a spectral phase kernel that emulates the birefringence and frequency channelized diffractive propagation. The final step of PAGE is to apply thresholding and morphological operations on the generated feature vectors in spatial domain to produce the final output. The PAGE output embeds the original image into a set of feature maps that select semantic information at different scale, orientation, and spatial frequency. These feature maps include spacial frequency bins and edge angle bins.

The spectral phase operator is expressed as a product of two phase functions, $\phi_1$ and $\phi_2$. The first component $\phi_1$ is a symmetric gaussian filter that selects the spatial frequency range of the edges that are detected. Default center frequency is 0, which indicates a baseband filter, the center frequency and bandwidth of which can be changed to probe edges with different sharpness — in other words to probe edges occurring over different spatial scales. The second component, $\phi_2$, performs the edge-detection. Since the output is based on the phase, it needs to be a complex-valued function. The PAGE operation transforms a real-value input to a complex-value quantity from which the phase is extracted.

This code is a simplified version of the full PAGE algorithm. It does not include the adaptive tone mapping operator introduced in [1], and it only includes one spatial frequency bin with a preselected bin central frequency, mu and bin bandwidth, sigma. It outputs a collection of angle dependent edges, each corresponding to one angle bin. For an N x M dimensional input, the output will be of size N x M x D, where D is the number of directional bins. This could be repeated for each color channel of a color image.

```
Parameters
----------
mu_1           # Center frequency of a normal/Gaussian distributed passband filter Phi_1
mu_2           # Center frequency of log-normal distributed distributed passband filter Phi_2
sigma_1        # Standard deviation sigma of normal/Gaussian distributed passband filter Phi_1
sigma_2        # Standard deviation sigma of log-normal distributed passband filter Phi_2
S1             # Strength (Amplitude) of Phi_1 filter
S2             # Strength (Amplitude) of Phi_2 filter
Direction_bins # Number of directional bins i.e. number of PAGE filter channels
sigma_LPF      # Standard deviation sigma of Gaussian distribution for smoothening kernel
Thresh_min     # Lower bound of bi-level (bipolar) feature thresholding for morphological operations
Thresh_max     # Upper bound of bi-level (bipolar) feature thresholding for morphological operations
Morph_flag     # Flag to choose (0) analog edge output or (1) binary edge output

import numpy as np
import mahotas as mh

# Define a function cart2pol that take in X, Y in cartesian and returns arrays theta, rho in polar coordinates
def cart2pol(x, y):
    theta = np.arctan2(y, x)
    rho = np.hypot(x, y)

    return (theta, rho)

def PAGE(I,handles):
    # Define two dimensional cartesian (rectangular) vectors, X and Y
    Image_orig_size=I.shape
    L=0.5
    u = np.linspace(-L, L, I.shape[0])
```

```python
v = np.linspace(-L, L, I.shape[1])
[U1, V1] = (np.meshgrid(u, v))
U = U1.T
V = V1.T

[THETA, RHO] = cart2pol(U, V)

# Low pass filter the original image to reduce noise
Image_orig_f = ((np.fft.fft2(I)))
expo = np.fft.fftshift(np.exp(-0.5*np.power((np.divide(RHO, np.sqrt((handles.sigma_LPF ** 2) / np.log(2)))), 2)))
Image_orig_filtered = np.real(np.fft.ifft2((np.multiply(Image_orig_f, expo))))

# Create PAGE directional filters
Min_Direction = 1 * np.pi / 180 # Direction in radians
Direction_span = np.pi / handles.Direction_bins # Direction step in radians
Direction = np.arange(Min_Direction, np.pi, Direction_span) # Direction array in radians

# Define the dimension of filter
X_size = Image_orig_size[0]
Y_size = Image_orig_size[1]
Z_size = Direction.shape[0]
PAGE_filter_array = np.zeros([X_size, Y_size, Z_size])

# For the number of directional bins, create PAGE kernels based on spectral directionality
for i in range(Z_size):
    tetav = Direction[i]

    # Project onto new directionality basis for PAGE filter creation
    Uprime = U * np.cos(tetav) + V * np.sin(tetav)
    Vprime = -U * np.sin(tetav) + V * np.cos(tetav)

    # Create Normal component of PAGE filter
    Phi_1 = np.exp(-0.5 * ((abs(Uprime) - handles.mu_1) / handles.sigma_1)** 2) / (
            1 * np.sqrt(2 * np.pi) * handles.sigma_1)
    Phi_1 = (Phi_1 / np.max(Phi_1[:])) * handles.S1

    # Create Log-Normal component of PAGE filter
    Phi_2 = np.exp(-0.5 * ((np.log(abs(Vprime)) - handles.mu_2) / handles.sigma_2) ** 2) / (
            abs(Vprime) * np.sqrt(2 * np.pi) * handles.sigma_2)
    Phi_2 = (Phi_2 / np.max(Phi_2[:])) * handles.S2

    # Add overall directional filter to PAGE filter array
    PAGE_filter_array[:,:,i]= Phi_1 * Phi_2  # log Normal distribution for X * Gaussian Distribution for Y

# Initialized PAGE output based on original image and number of directional bins
PAGE_output = np.zeros((Image_orig_filtered.shape[0],Image_orig_filtered.shape[1],Direction.shape[0]))

# Looping over number of directional bins, apply each PAGE directional kernel and save as channel of PAGE_output
for j in range(Direction.shape[0]):
    # Apply PAGE Kernel in spectral domain
    temp = (np.fft.fft2(Image_orig_filtered)) * np.fft.fftshift(np.exp(-1j * PAGE_filter_array[:,:,j]))
    Image_orig_filtered_PAGE = np.fft.ifft2(temp)

    # Take phase of output in spacial domain
    PAGE_Features = np.angle(Image_orig_filtered_PAGE)

    # Morphological operations: For Morph_flag == 0, return analog edges. For Morph_flag == 1, erode edges to binary.
    if handles.Morph_flag == 0:
        out = PAGE_Features
        PAGE_output[:,:,j] = out
```

```python
        else:
            features = np.zeros(PAGE_Features.shape);
            features[np.where(PAGE_Features > handles.Thresh_max)] = 1
            features[np.where(PAGE_Features < handles.Thresh_min)] = 1  # Output phase has both positive and negative values
            features[np.where(I < np.max(np.max(I)) / 20)] = 0  #  Ignore the features in the very dark areas of the image

            out = features
            out = mh.thin(out, 1)
            out = mh.bwperim(out, 4)
            out = mh.thin(out, 1)
            out = mh.erode(out, np.ones((1, 1)))
        PAGE_output[:,:,j]=out

    return PAGE_output, PAGE_filter_array
```